\documentclass[twocolumn]{aastex63}
\usepackage{graphics,epsf}
\usepackage{amsmath}                
\usepackage{amsfonts}               
\usepackage{amssymb}                
\usepackage{epsfig}                 
\usepackage{appendix}
\usepackage{graphicx}
\usepackage{float}
\usepackage{color}
\usepackage{multirow}
\usepackage{colortbl}
\usepackage[para,online,flushleft]{threeparttable}

\newcommand{\cm}{{~\rm cm}}

\newcommand{\km}{{~\rm km}}

\newcommand{\g}{{~\rm g}}

\newcommand{\erg}{{~\rm erg}}
\newcommand{\yr}{{~\rm yr}}

\begin{document}

\title{Astrophysical Naturalness}

\email{soker@physics.technion.ac.il}

\author{Noam Soker}
\affiliation{Department of Physics, Technion – Israel Institute of Technology, Haifa 3200003, Israel}

\begin{abstract}
I suggest that stars introduce mass and density scales that lead
to `naturalness' in the Universe. Namely, two ratios of order
unity. (1) The combination of the stellar mass scale,
$M_\ast(c,\hbar,G, m_p, m_e, e, \dots)$, with the Planck mass,
$M_{\rm Pl}$, and the Chandrasekhar mass leads to a ratio of order
unity that reads $N_{{\rm Pl \ast}} \equiv  {M_{\rm Pl}}/({M_\ast}
m^2_p )^{1/3} \simeq 0.15 - 3$, where $m_p$ is the proton mass.
(2) A system with a dynamical time equals to the nuclear life times of stars, $\tau_{\rm nuc \ast}$, has a density of $\rho_{\rm D \ast}(c,\hbar,G, m_p, m_e, e, \dots) \equiv ( G ~ \tau^2_{\rm nuc \ast} ) ^{-1}$. The ratio of the dark energy density to this density is $N_{\lambda \ast} = {\rho_{\Lambda}}/{\rho_{\rm D \ast}}   \approx 10^{-7} - 10^{5}$.
Although the range is large, it is critically much smaller than
the 123 orders of magnitude usually referred to when
$\rho_{\Lambda}$ is compered to the Planck density. In the pure
fundamental particles domain there is no naturalness; either
naturalness does not exist or there is a need for a new physics or
new particles. The `Astrophysical Naturalness' offers a third
possibility: stars introduce the combinations of, or relations
among, known fundamental quantities that lead to naturalness.
\end{abstract}



\section{Introduction}
\label{sec:MN}

The naturalness topic is nicely summarized by Natalie Wolchover in
an article from May 2013 in  Quanta
Magazine\footnote{https://www.quantamagazine.org/20130524-is-nature-unnatural/}.
I here discuss two points as listed in the talk ``Where are we
heading?'' given by Nathan Seiberg in
2013:\footnote{http://physicsforme.com/2013/07/24/nathan-seiberg-where-are-we-heading/}
(1) ``Why doesn't dimensional analysis work? All dimensionless
numbers should be of order one''; (2) ``The cosmological constant
is quartically divergent - it is fine tuned to 120 decimal
points.''

My answer to the first point is that in astrophysics dimensional
analysis does work when stars are considered as fundamental
entities. This answers the second question as well. If the nuclear
lifetime of stars is taken to be a dynamical time of the Universe,
then naturalness emerges from the observed cosmological constant.
No fine tuning is required.

Many relations among microscopic quantities and their relations
with macroscopic quantities are discussed by \cite{CarrRees1979}
who try to explain these relations, or else they refer to the
anthropic principle to account for some of the relations. A more
recent study was conducted by \cite{BurrowsOstriker2014}. Here I
do not repeat the explanations in those two papers. I simply take
stars to provide the relations among the many physical constants
and particles properties, and show that naturalness emerges from
the relations introduced by stars. One might refer to relations
introduced by stars as coincidental (e.g.,
\citealt{CarrRees1979}), but in the present essay I prefer to
refer to these relations as naturalness. My goal is to suggest a
third option to treat naturalness, as I explain in the last
section.

This essay does not discover anything new, but rather suggests to
include stars as `fundamental entities' when considering
naturalness in our Universe. As naturalness was discussed in talks
and popular articles, I use them as references. I also limit the
discussion to two commonly discussed quantities in relation to
naturalness, the Planck mass and the cosmological constant (dark
energy). Many other relations and coincidences can be found in \cite{CarrRees1979} and \cite{BurrowsOstriker2014}. I
will not touch the question of multiverse which is often connected
to the values of fundamental quantities (e.g., \citealt{Weinberg2005, 
LivioRees2005, LivioRees2018, Adams2019, AlonsoSerranoJannes2019}).

\section{The Chandrasekhar mass}
\label{sec:MbCh}

The Planck mass that starts the discussion on naturalness is
defined as
\begin{equation}
M_{\rm Pl}= \left( \frac{\hbar c}{G} \right)^{1/2} = 2.177 \times
10^{-5} \g.
  \label{eq:mpl}
\end{equation}
It is many orders of magnitude above the mass of the Higgs boson
and all other fundamental particles. If we constrain ourselves to
the particle world, no naturalness exists (e.g.
\citealt{Dine2015}). Let us add stars.

Consider the Chandrasekhar mass limit $M_{\rm Ch}$. This is
the maximum mass where a degenerate electron gas can support a
body against gravity. The electrons are relativistic at this mass
limit, and the expression reads
\begin{equation}
M_{\rm Ch}=K_1 \left( \frac{Z}{A} \right)^2
 \left(  \frac{M_{\rm Pl}}{m_p} \right)^2 M_{\rm Pl} = K_1 \left( \frac{Z}{A} \right)^2  M_{\rm BCh} ,
  \label{eq:mch}
\end{equation}
where $m_p=1.673 \times 10^{-24} \g$ is the proton mass, and  $Z$
and $A$ are the atomic number and atomic mass number,
respectively, of the element(s) composing the white dwarf (the
ratio $Z/A$ is the mean number of electrons per nucleon in the
white dwarf). The constant $K_1 \simeq 3.1$ is composed of pure
numbers (no physical constants), and $K_1 (Z/A)^2 \simeq 0.8$ for
white dwarfs in nature where $Z=0.5A$. The last equality defines
what I term the bare Chandrasekhar mass
 \begin{equation}
M_{\rm BCh} \equiv  \left(  \frac{M_{\rm Pl}}{m_p} \right)^2
M_{\rm Pl} = \alpha_G^{-1} M_{\rm Pl} = 1.85 M_\odot ,
  \label{eq:bmch}
\end{equation}
where $\alpha_G=Gm^2_p/\hbar c=5.9\times 10^{-39}$ is the
gravitational fine structure constant, which is also used to
express $M_{\rm BCh}$ (e.g., \citealt{CarrRees1979}).

\section{Naturalness with stars}
\label{sec:Nat}

The mass of stars, {{{{ namely, gravitationally bound objects that sustain hydrogen nuclear burning, }}}} is determined by the requirement that hydrogen
burns to helium. From below it is limited by brown dwarfs, where
the star cannot compress and heat enough to ignite hydrogen. The
minimum mass for a star is $M_\ast> 0.08 M_\odot$. The maximum
stellar mass of hundreds solar masses is not well determined, but
radiation pressure limits the upper mass (e.g.,
\citealt{CarrRees1979, BurrowsOstriker2014}). Interestingly, the
Chandrasekhar mass sits more or less in the center of the stellar
mass range in logarithmic scale (e.g., \citealt{CarrRees1979})
\begin{equation}
N_{{\rm M \ast}} \equiv \frac{M_{\rm BCh}}{M_\ast} = \left(
\frac{M_{\rm Pl}}{m_p} \right)^2 \frac{M_{\rm Pl}}{M_\ast} \simeq
0.01 - 20.
  \label{eq:nm}
\end{equation}
In the logarithmic scale the range of this ratio is approximately
$-2$ to $1.4$, much-much smaller than the 17 orders of magnitude
difference between the mass of the Higgs boson and the Planck
mass. Moreover, if the ratio is with the Planck mass rather than
$M_{\rm BCh}$, then the ratio is closer to unity, as it reads
\begin{equation}
N_{{\rm Pl \ast}} \equiv \frac{M_{\rm Pl}}{\left( M_\ast m^2_p
\right)^{1/3}}  \simeq 0.15 - 3.
  \label{eq:npl}
\end{equation}

It is important to emphasise that the mass of stars is determined
by the requirement that hydrogen experience thermonuclear burning
to helium. The Chandrasekhar mass is determined from the pressure
that a degenerate electrons gas can hold against gravity. Nothing
demands them to be equal. But they are. Namely,  the ratio of the
Chandrasekhar mass, that is composed of the Planck and the proton
masses, to stellar mass is of order one. \emph{This is naturalness.}

Of course, the properties of stars are determined by the
properties of the four fundamental forces, as all of them are
involved in the nuclear burning and stellar structure, and the
properties of the particles involved. The question is what
combination of the fundamental constants of the forces and of the
particles' properties gives two quantities whose ratio is $\approx
1$? The answer here is that stars form this combination as
 \begin{equation}
  M_\ast=M_\ast(c,\hbar,G, m_p, m_e, e, {\rm Forces~of~nature,}
  \dots),
  \label{eq:Mstarf}
 \end{equation}
 hence give us the naturalness in the Universe. This is expressed in
equations (\ref{eq:nm}) and (\ref{eq:npl}).
{{{{ The stellar structure is actually sensitive to $e^2$ more than to $e$ (the charge of the electron). With the constants $\hbar$ and $c$, one can rather write the fine structure constant as an independent variable in equation (\ref{eq:Mstarf}). }}}}

In other words, much as the proton `forms' a combination from the
properties of the quarks and the electric and color forces to give
a mass, the proton mass $m_p$, so do stars. But stars build a much
more complicated combination, and with many more of the
fundamental constants and forces, and the output of this relation
is not quantized, but it is rather a continuous function.

I note that \cite{CarrRees1979} try to show that $N_{{\rm Pl
\ast}} \sim 1$ is expected. However, they had to use numbers from
more complicated calculations than just order of magnitude
estimates. They specifically use the nuclear burning temperature
of hydrogen, $T_H$, and take a factor of  $q \sim 10^{-2}$ in the
expression $k T_H = q m_e c^2$. \cite{BurrowsOstriker2014}
consider the ratio between the maximum stellar mass and the
Chandrasekhar mass, and take the extra (external) factor that
comes from observations and detailed calculations to be the ideal
gas pressure to total pressure ratio $\beta$. In setting the lower
stellar mass limit \cite{BurrowsOstriker2014} take another extra
parameter to get the burning temperature of hydrogen. The
parameter is the ratio of the Gamow energy to $k_B T/3$, which
they set equal to 5. That is, it is not trivial to express stellar
properties from fundamental particles and physical constants. My
approach is different. I avoid these extra parameters. I take
stars to simply provide the relations among the different
quantities.

There is also the demand that the baryonic density in the Universe
be high enough for stars to form in the first place (e.g.,
\citealt{CarrRees1979, LivioRees2005}). A related natural ratio is
discussed in section \ref{sec:de}

{{{{ \cite{Chandrasekhar1978} in his article where he refers also to the work of Eddington, mentions that when the ratio of radiation pressure to total pressure $(1-\beta)$ is neither too close to zero nor too close to unity, namely $0.01 \la (1-\beta) \la 0.9$, then the stellar mass is of the order of the Chandrasekhar mass. This argument holds whether nuclear burning powers the star or whether gravitational contraction powers the star. 
\cite{Chandrasekhar1978} emphasises the amazing coincidence that the masses of radiating globes that result from Eddington’s argument are of the same order as the Chandrasekhar mass, and of the masses that are required to ignite nuclear burning in the interior.
 }}}}

\section{Stellar explosion energy}
\label{sec:Nuc}

The naturalness has several implications. One of them is that
regular stars can lead to white dwarfs with a mass close to and
above the Chandrasekhar mass. White dwarfs with that mass or
above, and iron cores of massive stars with that mass, explode
eventually as a supernova. White dwarfs explode as thermonuclear
supernovae where carbon and oxygen burn to nickel; cores of
massive stars explode as core-collapse supernovae where a neutron
star is formed. The typical kinetic energy of the ejected gas in
supernovae, $\approx 10^{51} \erg$, can be derived from
fundamental quantities.

 The radius of an idealized white dwarf supported by a degenerate
non-relativistic electrons gas is given by
\begin{equation}
R_{\rm WD}= K_2
  \frac {\hbar^2} {G ~m_e ~m^{5/3}_p ~M^{1/3}_{\rm WD}}
  \left( \frac{Z}{A}\right)^{5/3},
  \label{eq:rwd}
\end{equation}
where $M_{\rm WD}$ is the white dwarf mass and the constant $K_2
\approx 1$ is composed of pure numbers. For other forms of this
expression for the white dwarf radius see
\cite{BurrowsOstriker2014}. Although for a white dwarf at the
Chandrasekhar mass the electrons gas is degenerate, I nonetheless
substitute the bare Chandrasekhar mass $M_{\rm BCh}$ in equation
(\ref{eq:rwd}) to estimate for the bare white dwarf
radius
\begin{equation}
R_{\rm BWD} \equiv
  \frac {\hbar^2}{G ~m_e ~m^{5/3}_p ~M_{\rm BCh}^{1/3}}
   = \frac {G}{m_e c^2} \frac{M^3_{\rm Pl}}{m_p} = 5000 \km. 
  \label{eq:rmch}
\end{equation}
Due to the factor $(Z/A)^{5/3}$ the real radius is smaller by a
factor of $\approx 3$. We can define the bare gravitational-energy
of the bare white dwarf as
\begin{equation}
E_{\rm BCh} \equiv \frac{G M^2_{\rm BCh}}{R_{\rm BWD}} =
    \left( \frac{M_{\rm Pl}}{m_p} \right)^3 m_e c^2
     =  1.8 \times 10^{51} \erg .
  \label{eq:Emch}
\end{equation}
This is the typical kinetic energy of the mass ejected in
supernova explosions of either massive stars (core collapse
supernovae) or of white dwarfs (Type Ia supernovae). Simply the
explosion energy is of the order of the binding energy of an electron-degenerate star.

Accurate calculations give lower binding energy values to
exploding white dwarfs and collapsing cores by a factor of
several. This is because the internal energy has a positive value.
The explosion kinetic energy is then several times the binding
energy of the degenerate core. But this does not change the argument.

The factor $( {M_{\rm Pl}}/{m_p} )^3$ is the number of nucleons in
the white dwarf, so that the binding energy per nucleon is
$\approx m_e c^2$. This is also the order of magnitude of the
nuclear energy released per nucleon when carbon and oxygen burn to
nickel. This nuclear energy is the energy source of type Ia
supernovae. Indeed, about $20-60 \%$ of the white dwarf burns to
nickel during a type Ia supernova.

When a core of a massive star collapses to a neutron star it
releases a total energy of $\approx {\rm few} \times 10^{53}
\erg$. This energy comes from the final radius of the neutron star
which is determined from nuclear repulsive forces acting against
gravity. Most of this energy is carried out by neutrinos (and
anti-neutrinos) of the three kinds. 

\section{The naturalness of dark energy}
\label{sec:de}

The usual approach to search for naturalness is to compare the
observed density of the dark energy $\rho_{\Lambda}= 7 \times
10^{-30} \g \cm^{-3}$ with the Planck density $\rho_{\rm
Pl}(c,\hbar,G)=M_{\rm Pl} l^{-3}_{\rm Pl}= 5.155\times 10^{93} \g
\cm^{-3}$, where $l_{\rm Pl}=(\hbar G/c^3)^{1/2}$ is the Planck
length. This gives an `unnatural' ratio of
$U=\rho_{\rm Pl}(c,\hbar,G)/\rho_{\Lambda}= 10^{123}$. We in
astrophysics are not accustomed to such astronomical numbers. This
`unnatural' ratio is referred to as the cosmological constant
problem (e.g., \citealt{Carroll2002}). A density function to
replace $M_{\rm Pl} l^{-3}_{\rm Pl}$ is required.

As we saw in previous sections, stars introduce a (complicated)
combination of the fundamental quantities to give a mass ratio of
order unity, that is, an astrophysical mass naturalness (equations
\ref{eq:nm} and \ref{eq:npl}). Stars also introduce some typical time scales, like their dynamical time scale, thermal time scale, and nuclear life
time. I take here the nuclear time scale which is the life time
over which a star evolves, as I compare the quantity with the dark
energy that is related to the evolution of the Universe.

Stars spend most of their nuclear lives burning hydrogen to
helium. The nuclear life time of stars depends mainly on the
initial mass of the star, with $\tau_{\rm nuc \ast}(0.1M_\odot)
\approx 10^{13} \yr$, $\tau_{\rm nuc \ast}(M_{\rm BCh}) \approx
10^{9} \yr$, and $\tau_{\rm nuc \ast}(M > 10M_\odot) \approx
10^{7} \yr$. 

{{{{ I ask now the following question. What system will have a dynamical time scale $t_{D\ast}$ that is equal to the nuclear life-time of stars $\tau_{\rm nuc}$? The answer is a system that has an average density of }}}}
\begin{equation}
\rho_{\rm D \ast} \equiv \left( G ~ \tau^2_{\rm nuc \ast} \right)
^{-1} \approx 10^{-34} - 10^{-22} \g \cm^{-3}.
  \label{eq:rhod}
\end{equation}
This density comes from the nuclear life time of stars that depends
on many fundamental parameters. Namely,
 \begin{equation}
  \rho_{\rm D \ast}=\rho_{\rm D \ast}(c,\hbar,G, m_p, m_e, e, {\rm Forces~of~nature,} \dots) .
  \label{eq:rhodf}
 \end{equation}
The point here is that stars introduce the basic relation among
these fundamental quantities.

The second natural number defined in this essay is therefore
\begin{equation}
N_{\lambda \ast} \equiv \frac{\rho_{\Lambda}}{\rho_{\rm D \ast}}
\approx 10^{-7} - 10^{5}.
 \label{eq:nlambda}
\end{equation}
Although the range is large, it is critically much smaller than
the 123 orders of magnitude usually referred to when
$\rho_{\Lambda}$ is compered to `natural density'. Moreover, a
ratio of unity sits just near the center of this range.

I conclude that stars introduce a nuclear time scale, whose
associated dynamical time scale leads to a density about equal to
the dark energy density. Again, the nuclear time scale of stars is
determined by a complicated relation of fundamental quantities,
constants and particle properties. Stars combine the fundamental
quantities to lead to a naturalness.

It is important to emphasise that the approach here is different
than the question ``Why did the cosmological constant (dark
energy) become significant only recently? (e.g.,
\citealt{LivioRees2005}). Namely, why the age of the universe is
about equal to the dynamical time associated with the density of
the cosmological constant? The approach here also differs from
coincidental identities that are related to the present age or
size of the Universe (e.g., \citealt{CarrRees1979}).

In the present approach the age of the universe has no importance
at all. The same argument presented here holds as soon as hydrogen
becomes the main element in the universe; the first minute of the
universe, at an age of $10^{-16}$ times the present Universe age.
The same argument will be true when the universe be $10^{16}$
times its present age (as long as the dark energy density stays
constant; see section \ref{sec:TimeVariation}).

It is true that if the cosmological constant (dark energy) had
been much larger, stars would not have formed (see, e.g.,
\citealt{Garrigaetal2000}, and also for other time scales
involving the cosmological constant). The value of primordial
density fluctuations is also related to the question of star
formation \citep{LivioRees2005}. But here I don't examine these
questions; I look for ratios of order unity, i.e., naturalness.

\section{Implications on variation of the fundamental constants}
\label{sec:TimeVariation}
 
The astrophysical naturalness approach disfavours any time-variation of the fundamental constants of nature. In principle, the different constants can vary as to maintain the ratios (\ref{eq:nm}), (\ref{eq:npl}), and (\ref{eq:nlambda}) around unity. However, the typical stellar mass $M_\ast$ (eq. \ref{eq:Mstarf}) and the density that the nuclear life time of stars introduces $\rho_{\rm D \ast}$ (eq. \ref{eq:rhod}) depend in a very complicated manner on many fundamental constants. It will require a fine-tuned evolution with time of the different constant to maintain theses ratios at about unity.  
 
This holds as well to the value of the cosmological constant $\lambda$. Namely, the astrophysical naturalness that section \ref{sec:de} presents requires, if we to avoid fine tuning, that the cosmological constant is indeed constant and does not vary with time.  

Overall, the astrophysical naturalness approach, that holds that stars, despite being very complicated, serve as a basic entity in our Universe, makes the Universe simpler in both introducing naturalness and in arguing that fundamental constants, including the cosmological constant (dark energy), do not vary with time.

\section{Summary}
\label{sec:expe}

The naturalness question I studied here can be posed as follows:
``What is the combination of the fundamental constants and
particle properties that leads to a ratio of two values that is of
order unity?'' In the present essay I showed that stars introduce
these combinations that give what might be termed ``Astrophysical
Naturalness''.

Stars introduce the stellar mass given in equation
(\ref{eq:Mstarf}) that leads to the natural relation
(\ref{eq:nm}), or (\ref{eq:npl}). Stars also introduce a nuclear
timescale. If this time scale is associated with a dynamical time
scale, then a density $\rho_{\rm D \ast}$ given by a very
complicated relation (eq. \ref{eq:rhodf}) is defined. This density
leads to the natural relation (\ref{eq:nlambda}).

Nathan Seiberg summarizes his talk by a diagram that leaves two
basic options, ($i$) abandon naturalness, or ($ii$) go beyond
known physics/particles to find naturalness. Here I take a third
option which is basically to add stars as a basic entity in our
Universe, much as the proton is a composite particle. This brings
out naturalness in a beautiful way, at least in the eyes of an
astrophysicist.

In section \ref{sec:TimeVariation} I argued that the astrophysical naturalness approach suggests that fundamental constants, including the cosmological constant (dark energy), do not vary with time.

The arguments presented here are not the anthropic principle,
e.g., as presented by \cite{LivioRees2005}. \cite{LivioRees2005}
list the necessity of stars to form in order to have life. I
differ here in two respects. (1) I treat stars on the same level
as I treat baryons. I do not require that the properties of
protons allow stars as \cite{LivioRees2005} do. I simply treat
stars as I treat baryons (although definitely stars are more
complicated and composed of baryons). Both baryons and stars are
composite entities that exist in the Universe. They appear on the
same level in equations (\ref{eq:nm}) and (\ref{eq:npl}). (2) The
arguments presented here do not require the presence of carbon in
the universe. All arguments here apply if nuclear reactions would
have ended with helium. As well, life requires some chemical
properties. The arguments presented here don't involve chemistry
at all. For example, even if all stars were much hotter and the
strong UV radiation would prevent life, the arguments presented
here still hold.

An overall summary is that the astrophysical approach makes the Universe simpler in both introducing naturalness and in arguing that fundamental constants do not vary with time.

\acknowledgments{I thank Adi Nusser for useful discussions, and two referees for very useful comments.}

\label{lastpage}
\end{document}